\documentclass[aps, prd, 12pt, preprint, a4paper, nofootinbib, superscriptaddress]{revtex4-1}

\usepackage{bm} 
\usepackage{bbm} 
\usepackage{amsthm}
\usepackage{slashed}
\usepackage{amssymb}
\usepackage{amsmath}
\usepackage{mathrsfs}
\usepackage{amsfonts}
\usepackage{combelow}
\usepackage{mathtools}
\usepackage{hyperref}

\usepackage{color}

\theoremstyle{definition}

\usepackage{braket}

\usepackage{physics}

\usepackage{tensor}

\DeclareMathAlphabet{\mathbb}{U}{msb}{m}{n}

\newcommand{\rn}{\mathbb{R}}

\newcommand{\cn}{\mathbb{C}}

\newcommand{\df}{\coloneqq}

\newcommand{\td}[1]{\tilde{#1}}

\newcommand{\al}{\alpha}

\newcommand{\lag}{\mathscr{L}}

\newcommand{\hp}{\hat{\pi} }

\newcommand{\heta
}{\hat{\eta}}

\newcommand{\hph}{\hat{\phi}}

\newcommand{\hu}{\hat{U}}

\newcommand{\hpr}{\hat{\mathcal{P}}}

\newcommand{\htr}{\hat{\mathcal{T}}}

\newcommand{\thph}[3]{\hat{\td{\phi}}^{#1}_{#2} (#3)}

\newcommand{\hh}{\hat{H}}

\newcommand{\hlag}{\hat{\lag}}

\newcommand{\vx}{\vec{x}}

\newcommand{\vy}{\vec{y}}

\newcommand{\intax}[1]{\int_{t_0,\:\vec{#1}}^{x^0}\:}

\newcommand{\intay}[1]{\int_{t_0,\:\vec{#1}}^{y^0}\:}

\newcommand{\be}{\begin{equation}}

\newcommand{\ee}{\end{equation}}

\newcommand{\bi}{\begin{itemize}}

\newcommand{\ei}{\end{itemize}}

\newcommand{\comma}{`}

\begin{document}

\setlength{\abovedisplayskip}{12pt}
\setlength{\belowdisplayskip}{12pt}

\title{Pseudo-real quantum fields}

\author{Maxim N. Chernodub}
\email{maxim.chernodub@univ-tours.fr}
\affiliation{Universit\'e de Tours, Universit\'e d'Orl\'eans, CNRS, Institut Denis Poisson, UMR 7013, Tours, 37200, France}
\affiliation{Department of Physics, West University of Timi\cb{s}oara,
Bulevardul Vasile P\^arvan 4, Timi\cb{s}oara 300223, Romania}
\author{Peter Millington}
\email{peter.millington@manchester.ac.uk}
\affiliation{Department of Physics and Astronomy, University of Manchester, Manchester M13 9PL, United Kingdom}
\author{Esra Sablevice}
\email{esra.sablevice@manchester.ac.uk}
\affiliation{Department of Physics and Astronomy, University of Manchester, Manchester M13 9PL, United Kingdom}

\date{September 18, 2025}

\begin{abstract}
We introduce the concept of pseudo-reality for complex numbers. We show that this concept, applied to quantum fields, provides a unifying framework for two distinct approaches to pseudo-Hermitian quantum field theories. The first approach stems from analytically continuing Hermitian theories into the complex plane, while the second is based on constructing them from first principles. The pseudo-reality condition for bosonic fields resolves a long-standing problem with the formulation of gauge theories involving pseudo-Hermitian currents, sheds new light on the resolution of the so-called Hermiticity Puzzle, and may allow a consistent minimal coupling of pseudo-Hermitian quantum field theories to gravity. We focus on the $i\phi^3$ cubic scalar theory, obtaining the relevant pseudo-reality conditions up to quadratic order in the coupling; a theory of two complex scalar fields with non-Hermitian mass mixing; and the latter's coupling to a $U(1)$ gauge field. The general principle of pseudo-reality, however, is expected to contribute to the ongoing development of the first-principles construction of pseudo-Hermitian quantum field theories, including their formulation in curved spacetimes.
~
\\
\footnotesize{This is an author-prepared post-print of \href{https://doi.org/10.1103/kn9t-f8vx}{Phys.\ Rev.\ D {\bf 112} (2025) 065007}, published by the American Physical Society under the terms of the \href{https://creativecommons.org/licenses/by/4.0/}{CC BY 4.0} license (funded by SCOAP\textsuperscript{3}).}
\end{abstract}

\maketitle

\newpage

\tableofcontents

\newpage


\section{Introduction}

The success of non-Hermitian quantum mechanics has led to increasing interest in extending these ideas to non-Hermitian quantum field theory, with much of the progress so far focused on $PT$-symmetric quantum field theories, where the Hamiltonian is symmetric under the combined action of parity $P$ and time-reversal $T$. When the $PT$ symmetry is exact, the Hamiltonian has a real spectrum and the theory follows a (pseudo-)unitary evolution, which ensures its physical viability~\cite{Bender:1, Bender:2, Bender:2009mq, Most:1, Most:2, Most:3}. Well-studied examples include scalar field theories with an imaginary cubic $i\phi^3$~\cite{Bphi1, Bphi2, Bphi3, Bphi4} or ``wrong-sign'' $-\phi^4$ quartic self-interaction~\cite{Sphi1, Sphi2, Bphi5}. 

However, non-Hermitian quantum field theories bring new challenges that are not present in non-Hermitian quantum mechanics. One such issue is the Hermiticity Puzzle~\cite{Mannheim:2018dur}, which arises when a Hermitian theory with real fields is made non-Hermitian by introducing complex couplings. This modification leads to equations of motion that are inconsistent under complex conjugation. Previous resolutions of the Hermiticity Puzzle required real fields to transform non-trivially under complex conjugation~\cite{Mannheim:2018dur}.

Inconsistent Euler-Lagrange equations are a common feature of non-Hermitian quantum field theories, as first noted in Ref.~\cite{Alexandre:2017foi}, and are not limited to theories with real fields. These theories also lack Poincaré invariance, and it remains unclear how to construct consistent gauge theories or couple them to gravity. In our previous work~\cite{Sablevice:2023odu} (see also Refs.~\cite{Alexandre:2020gah, Alexandre:2022uns,Chernodub:2021waz}), we traced these problems to a fundamental mismatch:\ the field $\phi$ and its Hermitian conjugate $\phi^\dag$ transform under different representations of the proper Poincaré group. This mismatch not only breaks Poincaré invariance but also causes $\phi$ and $\phi^\dag$ to evolve under different Hamiltonians, $H$ and $H^\dag$, respectively. To address this, we proposed replacing the Hermitian conjugate field $\phi^\dag$ with a dual field $\tilde{\phi}^\dag$, which transforms under the same representation of the proper Poincar\'e group as $\phi$ and thus evolves with the same Hamiltonian.

In this paper, we extend this idea to resolve the Hermiticity Puzzle by introducing the concept of pseudo-reality. Fields that are real in Hermitian theories become pseudo-real in pseudo-Hermitian theories, i.e., rather than being equal to their Hermitian conjugates $\phi=\phi^\dag\in\mathbb{R}$, they are equal to their duals $\phi=\td{\phi}^\dag\in \mathbb{C}$. Thus, the fields maintain the same number of degrees of freedom as a real scalar field, while being complex. This approach resolves a long-standing question about how to gauge non-Hermitian quantum field theories and may allow to couple them to gravity in a way that is consistent with the local symmetries of the Lagrangian and current conservation. Additionally, it unifies two approaches to constructing non-Hermitian quantum field theories found in the literature:\ one based on the analytic continuation of Hermitian theories into the complex space, and the other based on building the non-Hermitian theory from first principles.

The paper is organised as follows. In Section 2, we introduce the concept of pseudo-reality for complex functions and establish the conditions that a function must satisfy to be pseudo-real, providing examples for illustration. In Section 3, we apply the pseudo-real formalism to the non-Hermitian \(i\phi^3\) quantum field theory. We demonstrate that not only must the field be complex, but the parity and time-reversal transformations must also be non-trivial. We find the transformation under which the Hamiltonian is pseudo-Hermitian. In Section 4, we extend this approach to higher-spin fields. We show how to consistently gauge a two complex scalar model with non-Hermitian mass mixing and conjecture how to couple this model to gravity.


\section{Definition of pseudo-reality for complex numbers}
\label{sec:pseudo_real_functions}

The number $z$ is said to be real if it satisfies the following condition:
\begin{align}
    z = {\bar z} \qquad\ {\text{[for a real $z$]}}\,.
    \label{eq_reality}
\end{align}

The standard definition of reality~\eqref{eq_reality} for a complex number $z$ involves only the operation of complex conjugation $z \to {\bar z}$. The notion of reality can be generalized if one considers the reality condition with respect to a complex function $f = f (z)$. To this end, we consider a complex map $f:\,\mathbb{C}\mapsto\mathbb{C}$. If the function $f=f(z)$ is analytic and satisfies the property
\begin{align}
    f^{(-1)}(z) = {\bar f}(z)\,,
    \label{eq_f_pseudo_real}
\end{align}
valid for a continuous segment of the arguments $z\in\mathbb{C}$, where $\bar{f}$ is the complex conjugate of the map $f$, then $f$ can be used to define a curve in the complex plane via
\begin{align}
    z = f({\bar z})\,.
    \label{eq_z_pseudo_real}
\end{align}

We then call the number $z$ a pseudo-real number with respect to the complex map $f$ ($z\in\mathbb{R}_f$), provided the pseudo-reality condition~\eqref{eq_z_pseudo_real} is fulfilled and the function $f$ satisfies the consistency condition~\eqref{eq_f_pseudo_real}. Notice that Eq.~\eqref{eq_z_pseudo_real} implies automatically that
\begin{equation}
    \bar{z}=f^{(-1)}(z)\;,
    \label{eq_bar_z_pseudo_real}
\end{equation}
which can also be used as a definition of the pseudo-reality, fully consistent with Eq.~\eqref{eq_z_pseudo_real}.

Condition~\eqref{eq_f_pseudo_real} involves the inverse function $f^{(-1)}$ defined via $f^{(-1)}\Bigl(f(z)\Bigr) = z$. Therefore, the complex conjugation on the right-hand side of Eq.~\eqref{eq_f_pseudo_real} acts only on the function $f$, excluding its argument $z$, i.e.,
\begin{align}
    {\bar f}(z) \equiv \overline{f({\bar z})}\,.
\end{align}
Thus, Eq.~\eqref{eq_f_pseudo_real} can be understood in terms of the Laurent series of the function $f$ and its inverse $f^{(-1)}$ as follows:
\begin{align}
    {\rm if}\quad f(z) = \sum_{n \in {\mathbb Z}} C_n z^n
    \quad {\rm and} \quad
    f^{(-1)}(z) = \sum_{n \in {\mathbb Z}} C^{(-1)}_n z^n
    \quad {\rm then} \quad {\overline C}_n = C^{(-1)}_n, \ \ \forall n \in {\mathbb Z}\,.
\end{align}

If one writes $z = x + i y$, then Eq.~\eqref{eq_z_pseudo_real} gives us the following constraints on the variables $x$ and $y$:
\begin{align}
    x = u(x,y)\,, \qquad y = v(x,y)\,,
    \label{eq_two_eqs}
\end{align}
where $f(\bar z) = u(x,y) + i v(x,y)$. The system of two equations~\eqref{eq_two_eqs} provides us, in general, with a solution in terms of a list of isolated points. In order to obtain a line, this set of generally nonlinear equations should be degenerate so that these two equations should be satisfied by a single function $y = y(x)$. It is the constraint~\eqref{eq_f_pseudo_real} that makes the system~\eqref{eq_two_eqs} degenerate.

Let us consider now a few simple examples.
\begin{itemize}
    \item Take first a linear function $f(z) = C_0 + C_1 z$, where $C_0$ and $C_1$ are complex coefficients. In order for this function to be appropriate for the definition of the pseudo-reality of the complex number $z$~\eqref{eq_z_pseudo_real}, the function $f = f(z)$ must satisfy the consistency condition~\eqref{eq_f_pseudo_real}. The latter condition gives us the solution for the function $f$ in the following form:
\begin{align}
    f(z) = e^{2 i \theta} z\,, \qquad \theta \in {\mathbb R}\,,
    \label{eq_f_linear_example}
\end{align}
thus implying that $C_0 = 0$ and $C_1 = e^{2 i \theta}$, where $\theta \in [0, 2\pi)$ is a real number. Then the pseudo-reality condition~\eqref{eq_z_pseudo_real} reads as follows:
\begin{align}
	z= e^{2 i \theta} {\bar z}\,.
    \label{eq_reality_theta}
\end{align}

The standard reality condition~\eqref{eq_reality} corresponds to a pseudo-real number that satisfies~\eqref{eq_reality_theta} with $\theta = 0$. For a general angle variable $\theta$, the solution of the pseudo-reality condition~\eqref{eq_reality_theta}  defines a straight line in the complex plane, $z = \rho e^{i \theta}$ with $\rho \in \mathbb{R}$. Alternatively, in terms of $z = x + i y$, it gives the line $y = x \tan \theta$ with the inclination of the angle $\theta$ to the abscissa. While trivial, this gives an intuitive example of what is meant by a pseudo-real number:\ it is one with only one degree of freedom but a non-trivial complex conjugate.

\item For a non-linear function, let us consider a generic perturbative solution in terms of $f(z) = C_1 z + C_2 z^2$, with $|C_2 z| \ll |C_1|$. Solving Eq.~\eqref{eq_f_pseudo_real} for small arguments, one can demonstrate that the solution is:
\begin{align}
    f(z) = e^{2 i \theta} z \pm i g e^{3 i \theta} z^2 + O(g^2 z^3)\,, \qquad g \in {\mathbb R}\,, \quad |g z| \ll 1\,.
\end{align}
Functions of this form will arise in the case of quantum field theories with non-Hermitian interaction terms, as we will see in the next section in the context of the $i\phi^3$ theory.

\item Consider a generic M\"obius transformation:
\begin{align}
    f(z) = \frac{a z + b}{c z + d}\,,
    \label{eq_Moebius}
\end{align}
with the inverse
\begin{align}
    f^{-1}(z) = \frac{d z - b}{-c z + a}\,.
\end{align}
The condition~\eqref{eq_f_pseudo_real}, which defines the allowed function $f$, gives us the following set of degenerate constraints:
\begin{align}
    {\overline {\mathsf M}} = ({\rm det}\, {\mathsf M}) {\mathsf M}^{-1}\,, 
    \qquad 
    {\mathsf M} = \begin{pmatrix}
        a & b \\
        c & d 
    \end{pmatrix}\,,
\end{align}
or $a = \bar d$, $b = - \bar b$, and $c = - \bar c$. The constraint is resolved in the form
\begin{align}
        {\mathsf M} = \begin{pmatrix}
        a & i \kappa_1 \\
        i \kappa_2 & a 
    \end{pmatrix}\,, \qquad \kappa_{1,2} \in {\mathbb R}\,, \quad a \in {\mathbb C}\,.
    \label{eq_Moebius_coeff}
\end{align}
Thus, the  M\"obius function~\eqref{eq_Moebius} with the coefficients~\eqref{eq_Moebius_coeff} satisfies the requirement~\eqref{eq_f_pseudo_real} imposed on the function $f = f(z)$ to be used in the definition~\eqref{eq_f_pseudo_real} of the pseudo-reality for a complex number $z$. Then, the linear manifold identified by Eq.~\eqref{eq_z_pseudo_real} has the following explicit form:
\begin{align}
    (x - x_0)^2 + (y - y_0)^2 = r_0^2\,.
\end{align}
This formula describes a circle centered at the position $(x,y) = (x_0,y_0)$ with the radius $r_0$, related to the components of the matrix ${\mathsf M}$, Eq.~\eqref{eq_Moebius_coeff}, by the following equations:
\begin{align}
    x_0 = \frac{a_2}{\kappa_2}\,, 
    \qquad
    y_0 = - \frac{a_1}{\kappa_2}\,, 
    \qquad
    r_0^2 = x_0^2 + y^2_0 - \frac{\kappa_1}{\kappa_2}\,.
\end{align}
Notice that, in the region of parameter space with $r^2_0 < 0$, no solution exists.

\end{itemize}

In the sections that follow, we will apply this concept of pseudo-reality to quantum fields.


\section{Pseudo-real $i\phi^3$ scalar field theory}
\label{sec:iphi3_theory}

The \(i\phi^3\) theory, extensively studied by Bender and collaborators~\cite{Bphi1, Bphi2, Bphi3, Bphi4}, is the quantum field theory counterpart of the quantum mechanical model \(H = p^2 + ix^3\), for which the spectrum was shown to be entirely real~\cite{Bender:1, Bender:2012yv}. Unlike the Hermitian \(g\phi^3\) theory, which is unphysical due to its unbounded energy, the non-Hermitian \(ig\phi^3\) theory has a bounded energy spectrum that remains real under exact $PT$ symmetry.
  
In this section, we revisit the \(i\phi^3\) theory using the first-principles approach introduced in previous work~\cite{Sablevice:2023odu}. We show that extending the coupling \(g\mapsto ig \) into the complex plane requires the field \(\phi\) to be complex and the \comma standard' parity and time-reversal transformations lead to a contradiction. To resolve this, we formulate the theory in terms of the field $\phi$ and its dual $\tilde{\phi}^\dag$, both evolving with the same Hamiltonian. We apply the concept of pseudo-reality, where \(\phi\) is no longer real but a complex pseudo-real field, and we look for a \comma parity-like' transformation $\eta_P$ and a \comma time-reversal-like' transformation $\eta_T$ with respect to which the Hamiltonian is pseudo-Hermitian and $PT$ symmetric, making this a self-consistent formulation.

The Lagrangian of the \(i\phi^3\) theory is often written as
\begin{equation}
    \label{eq_scalar_J}
    \mathcal{L}(x)=\frac{1}{2}\partial_{\mu}\phi(x)\partial^{\mu}\phi(x)-\frac{1}{2}m^2\phi^2(x)-\frac{ig}{3!}\phi^3(x)\;,
\end{equation}
where $m^2>0$ and $g\in\mathbb{R}$ is some real coupling. In the \comma standard' approach, $\phi$ is considered to be a real scalar field, so that non-Hermiticity arises through the complex coupling $ig$, with
\be
 \mathcal{L}^\dag(x)=\frac{1}{2}\partial_{\mu}\phi(x)\partial^{\mu}\phi(x)-\frac{1}{2}m^2\phi^2(x)+\frac{ig}{3!}\phi^3(x)\;.
\ee

The non-Hermitian Lagrangian in Eq.~\eqref{eq_scalar_J} is $PT$ symmetric under the \comma standard' parity and time-reversal transformations, where $\phi$ transforms as a pseudo-scalar field:
\begin{subequations}
\begin{align}\label{eq_parity_standard}
P:& \:\:\:\:\phi(x) \longmapsto \phi^P(x_P)=-\phi (x)\:,\:\:\:\:\:\:\:\:\:\: \:\:\:\:\:\:\:\hpr\hph(x_P)\hpr^{-1}=-\hph (x)\:,  \\
T:&\:\: \:\:\phi(x)\longmapsto \phi^T (x_T)=\phi (x)\:,\:\:\:\:\:\:\:\:\:\:\:\:\:\:\:\: \:\:\:\:\:\htr\hph (x_T)\htr^{-1}=\hph(x)\:,\label{eq_time_standard}\\
PT:&\:\:\lag(x)\longmapsto \lag^{PT}(x_{PT})= \lag(x)\:,\:\:\:\:\:\:\:\:\:\:(\hpr\htr)\hlag(x_{PT})(\hpr\htr)^{-1}=\hlag(x)\:.
\end{align}
\end{subequations}
Here, the transformation of the $c$-number field $\phi$ is given on the left, and the transformation of the field operator $\hph$, indicated by a caret, on the right.

The existence of an anti-linear symmetry implies that the Lagrangian is pseudo-Hermitian \cite{Most:3, Most:4}, i.e., there exists some linear transformation $\eta$ that takes the Lagrangian to its Hermitian conjugate:
\be\label{eq_lag_ph}
\eta: \:\lag (x)\longmapsto \lag^\eta (x_\eta)=\lag^\dag(x)\:,\:\:\:\:\:\:\heta\hlag(x_\eta)\heta^{-1}=\hlag^\dag(x)\:.
\ee
Here, we take into account that $\eta:x\mapsto x_\eta$ could be a coordinate transformation (e.g., parity, $P:x\mapsto x_P$).

Indeed, the Lagrangian in Eq.~\eqref{eq_scalar_J} is pseudo-Hermitian with respect to the \comma standard' parity transformation in Eq.~\eqref{eq_parity_standard}:
\be
    P:\: \lag(x)\longmapsto \lag^{P}(x_P)=\lag^\dag(x)\:, \text{\:\:\:\:\:\:\:}\hpr \hlag(x_P)\hpr^{-1}=\hlag^\dag (x)\:.
    \label{eq_L_P_transformation}
\ee
It is also anti-pseudo-Hermitian with respect to the \comma standard' time-reversal transformation in Eq.~\eqref{eq_time_standard}:
\be
T:\: \lag(x)\longmapsto \lag^T(x_T)=\lag^\dag(x)\:,\:\:\:\:\:\:\:\:\: \htr \hlag (x_T)\htr^{-1}=\hlag^\dag (x)\:.
    \label{eq_L_T_transformation}
\ee
We call this anti-pseudo-Hermiticity due to the time-reversal operator being anti-linear~\cite{Most:3}. The corresponding Hamiltonian
\be\label{eq_Hamiltonian}
H=\int {\rm d}^3\vec{x}\:\qty( \frac{1}{2}\dot \phi^2(t,\vec{x})+\frac{1}{2}(\nabla \phi(t,\vec{x}))^2+\frac{1}{2}m^2\phi^2(t,\vec{x})+\frac{ig}{3!}\phi^3(t,\vec{x}))
\ee
is also parity-pseudo-Hermitian and time-reversal-anti-pseudo-Hermitian:
\begin{subequations}
\begin{align}\label{eq_H_pph}
    P:\:& H \longmapsto H^\dag\:,\text{\:\:\:\:\:\:}\:\:\: \hpr\hh\hpr^{-1}=\hh^\dag\:,\\
T:\:& H \longmapsto H^\dag\:, \text{\:\:\:\:\:\:\:} \:\:\htr\hh\htr^{-1}=\hh^\dag\:.\label{eq_H_tph}
\end{align}
\end{subequations}


\subsection{Issues with the standard approach}
\label{eq:issues}

Despite its apparent consistency, the \comma standard' approach described above has several issues. First, the field $\phi$ cannot be a real scalar field. This can be seen directly from the equations of motion
\begin{subequations}
\begin{align}
    \Box_x\phi(x)+m^2\phi(x)+\frac{ig}{2}\phi^2(x)=0\;,\\ \Box_x\phi^{\dag}(x)+m^2\phi^{\dag}(x)-\frac{ig}{2}\phi^{\dag 2}(x)=0\;,
\end{align}
\end{subequations}
which are not compatible under Hermitian conjugation if we take $\phi=\phi^{\dag}$. Indeed, if we solve these equations perturbatively in $g$, we see that $\phi$ is complex at  order $g$: 
\be\label{eq_phi_expansion}
\phi (x)=\phi_{0}(x)+\frac{ig}{2}\int {\rm d}^4y\: G_{0}(x,y)\phi_{0}^2(y)+\mathcal{O}(g^2)\:,
\ee
where $\phi_{0}$ is the free real Klein-Gordon scalar field and $G_{0}(x,y)$ is the Green's function of the free Klein-Gordon operator, i.e., 
\begin{align}\label{eq_green}
\qty(\Box_x+m^2)\phi_{0}(x)&=0\;,\\
(\Box_x+m^2)G_{0}(x,y)&=-\delta^{(4)}(x-y)\;.
\end{align} 
We can also look at Hamilton's equations of motion for the $c$-number field and the field operator:
 \begin{align}\label{eq:Hamiltons_eqs}
    \qty{\phi(x),H}_{\rm PB}=\partial_t \phi (x)&\text{\:\:\:and\:\:\:}[\hph(x),\hh]=i\partial_t \hph (x)\:,
\end{align}
where \comma \comma PB'' is the Poisson bracket. By taking the Hermitian conjugates of Eq.~\eqref{eq:Hamiltons_eqs}, we see that $\phi^\dag$ does not evolve with the same Hamiltonian as $\phi$, since $H^\dag \neq H$, i.e.,
\begin{align}
    \qty{\phi^\dag(x),H^\dag}_{\rm PB}=\partial_t \phi^\dag (x)&\text{\:\:\:and\:\:\:}[\hph^\dag(x),\hh^\dag]=i\partial_t \hph^\dag (x)\:.
    \end{align}
Hence, for any canonical formalism with a non-Hermitian Hamiltonian, the fields must be complex. Moreover, for a pseudo-Hermitian theory, the Hamiltonian $\hat{H}$ and its Hermitian conjugate $\hat{H}^{\dag}$ do not commute, since the left- and right-eigenvectors of a non-Hermitian Hamiltonian are distinct. It then follows that the field $\hat{\phi}$ and its Hermitian conjugate $\hat{\phi}^{\dag}$ do not commute at equal times (except at the initial time surface) by virtue of their respective evolutions with $\hat{H}$ and $\hat{H}^{\dag}$.

The second issue arises from assuming the \comma standard' parity and time-reversal transformations of the field $\phi$ in Eqs.~\eqref{eq_parity_standard} and~\eqref{eq_time_standard}.
For example, transforming the Hamilton's equation~\eqref{eq:Hamiltons_eqs} under parity and using that the Hamiltonian is parity-pseudo-Hermitian~\eqref{eq_H_pph}, we have
\begin{align}
    \hpr\qty[\hph(x),\hh]\hpr^{-1}=\hpr\: i\partial_t \hph (x)\: \hpr^{-1} \longmapsto \qty[\hph(x),\hh^\dag]=i\partial_t \hph (x)\:,
\end{align}
which leaves us with a contradiction, because, as argued above, $\hph$ and $\hph^\dag$ do not commute at equal times. The same contradiction is also found under the action of the \comma standard' time-reversal transformation. Hence, the parity and time-reversal transformations which make the theory pseudo-Hermitian and $PT$ symmetric are not the \comma standard' transformations but instead are non-trivial (see Refs.~\cite{Alexandre:2020gah, Alexandre:2022uns, Sablevice:2023odu}). We find them explicitly in the next Section~\ref{sec:PandTtransform}.


\subsection{Pseudo-Hermitian formalism}

For a quantum field theory with a non-Hermitian Hamiltonian, the field $\phi$ and its Hermitian conjugate $\phi^\dag$ transform under different representations of the proper Poincar\'e group~\cite{Sablevice:2023odu}. Thus, not only do they not evolve with the same Hamiltonian, but also the Lagrangian composed of $\phi$ and $\phi^\dag$ is not Poincar\'e invariant. Instead, in Ref.~\cite{Sablevice:2023odu} (see also Refs.~\cite{Alexandre:2020gah, Alexandre:2022uns,Chernodub:2021waz}), we proposed to build the Lagrangian from the field $\phi$ and its \comma dual' field $\tilde{\phi}^\dag$. We 
 define the \comma dual' field $\tilde{\phi}^\dag$ as the field transforming under  the same representation of the proper Poincar\'e group as $\phi$. 

Given a Lagrangian $\hlag(x)$ that is pseudo-Hermitian with respect to some Hermitian operator $\heta=\heta^{\dag}$, i.e., $\heta\hlag (x_\eta)\heta^{-1}=\hlag^\dag(x)$, the \comma dual' field operator $\hat{\td{\phi}}^\dag$, in general, has a form~\cite{Sablevice:2023odu}
\be\label{eq_dual_operator}
\thph{\dag}{}{x}=\heta^{-1}\hph^\dag (x_\eta)\heta\: \pi\:.
\ee
Here, $x_\eta$ is the coordinate transformed under $\heta$ (e.g., parity $x_P$). For an $n$-component field $\hph^a$, where $a\in\{1,\cdots,n\}$, $\pi$ is a Hermitian  $n\times n$ matrix. Hence, for a $1$-component scalar field, $\pi\in\rn$ is some real constant to be determined. We can check that the \comma dual' field operator in Eq.~\eqref{eq_dual_operator} evolves with the same Hamiltonian as $\hat{\phi}$:
\be\label{eq_motion_dual}
\qty[\thph{\dag}{}{x},\hh]=i\partial_t \thph{\dag}{}{x}\:
\ee
and, for the Lagrangian~\eqref{eq_scalar_J}, it has the same equation of motion as $\hat{\phi}$; namely,
\be
\ \Box_x\thph{\dag}{}{x}+m^2\thph{\dag}{}{x} +\frac{ig}{2}\thph{\dag^2}{}{x}=0\:.
\ee

The corresponding \comma dual' $\tilde{\phi}^\dag$ of the $c$-number field $\phi$ can be defined via the map $\eta$~\eqref{eq_lag_ph}:
\be\label{eq_dual_function}
\tilde{\phi}(x)\df \pi \eta [\phi (x)]\:\implies\:\tilde{\phi}^\dag(x)=\eta^\dag [\phi^\dag (x)]\pi\:,
\ee
so that the map $\eta$ is related to the operator $\heta$ via
\be\label{eq_eta_operator_eta}
\eta [\hph(x)]=\heta \hph (x_\eta)\heta^{-1}\:.
\ee
We can take Eq.~\eqref{eq_eta_operator_eta} as a definition of the functional map $\eta$. This map is linear, and the Hermiticity of the operator $\heta$ implies that the transformation  satisfies $\eta^\dag=\eta^{-1}$, which is consistent with providing a degenerate system of constraints that determine a contour in the complex plane (see Section~\ref{sec:pseudo_real_functions}).

Thus, in the remainder of this paper, we study the $i\phi^3$ theory in Eq.~\eqref{eq_scalar_J} as a theory of a complex scalar field $\phi$ and its \comma dual' field $\tilde{\phi}^\dag$. However, a complex scalar field has two propagating degrees of freedom rather than one degree of freedom as described by the original real scalar field. In the following subsection, we argue that the field $\phi$, while being complex, maintains only one degree of freedom. We do this by replacing \comma real' numbers with \comma pseudo-real' numbers.


\subsection{Pseudo-reality}

To this end, we propose that, for a pseudo-Hermitian Hamiltonian, a \comma real' field is not equal to its Hermitian conjugate, i.e., $\phi^\dag\neq\phi$. Instead, it is equal to its \comma dual' field~\eqref{eq_dual_operator}, making it \comma pseudo-real'; specifically,
\be\label{eq_pseudo_reality_def}
\tilde{\phi}^\dag (x)=\phi (x)\;,\qquad \hat{\tilde{\phi}}^\dag (x)=\hat{\phi} (x)\:.
\ee
This constraint is consistent with the Hamiltonian being $\heta$-pseudo-Hermitian, i.e., $\hh^\dag=\heta \hh \heta^{-1}$, and the field operator $\hat{\phi}$, while being complex, describes the same number of degrees of freedom as the original real scalar field. 

The pseudo-reality condition gives a non-trivial relationship between $\hat{\phi}$ and its Hermitian conjugate $\hat{\phi}^\dag$:
\be
\label{eq_pseudo_reality}
\hph^\dag (x)= \pi \heta \hph (x_\eta)\heta^{-1}=\pi\eta[\hph (x)]\:,
\ee
where we have used Eq.~\eqref{eq_eta_operator_eta} relating the map $\eta$ to the operator $\heta$. 

Equation~\eqref{eq_pseudo_reality} presents a consistent resolution of the so-called Hermiticity Puzzle. Previous resolutions of this problem~\cite{Mannheim:2018dur} have required a redefinition of Hermitian conjugation, where real scalar fields, such as $\phi$ here, are odd under Hermitian conjugation, i.e., anti-Hermitian. In the present context, we see that this redefinition of Hermitian conjugation amounts to associating it with the composition of Hermitian conjugation and $\eta$-conjugation. 


\subsection{Parity and time-reversal}
\label{sec:PandTtransform}

Let us now assume that the field $\phi$ in the non-Hermitian Lagrangian~\eqref{eq_scalar_J} is complex and pseudo-real. This Lagrangian still has an anti-linear symmetry and is pseudo-Hermitian, but
not with respect to the ‘standard’ parity and time-reversal transformations in Eqs.~\eqref{eq_parity_standard} and~\eqref{eq_time_standard}. Instead, we have a linear \comma parity-like' and an anti-linear \comma time-reversal-like' transformation, which map the field $\phi$ to its \comma dual' conjugate $\tilde{\phi}$. For the \comma parity-like' transformation $\eta_P$, we have
\begin{subequations}
\begin{align}\label{eq_correct_parity}
    \eta_{P}:\:&\:\: \phi (x)\longmapsto \phi^{\eta_P}(x_P)=\eta_P [\phi (x)]=-\tilde{\phi} (x)\:,\:\:\:\:\:\qquad\qquad\heta_{P} \hph (x_P)\heta_{P}^{-1}=-\thph{}{}{x}\:,\\
&\lag(x)\longmapsto \lag^{\eta_P}(x_P)=\lag^\dag (x)\:,\qquad\qquad\qquad\qquad\:\:\:\: \heta_{P}\hlag (x_P)\heta_{P}^{-1}=\hlag^\dag (x)\:,
\end{align}
\end{subequations}
where the constant $\pi_P=-1$ in Eq.~\eqref{eq_dual_operator} reflects the pseudo-scalar nature of the field. For the \comma time-reversal-like' transformation $\eta_T$, we have
\begin{subequations}
\begin{align}\label{eq_correct_time}
    \eta_{T}:\:&\:\: \phi (x)\longmapsto \phi^{\eta_T}(x_T)=\eta_T[\phi(x)]=\tilde{\phi} (x)\:,\qquad\qquad\:\:\:\:\:\:\:\:\heta_{T} \hph (x_T)\heta_{T}^{-1}=\thph{}{}{x}\:,\\
&\lag(x)\longmapsto \lag^{\eta_T}(x_T)=\lag^\dag (x)\:,\qquad\qquad\qquad\qquad\:\:\:\: \heta_{T}\hlag (x_T)\heta_{T}^{-1}=\hlag^\dag (x)\:,
\end{align}
\end{subequations}
with  $\pi_T=+1$, as the anti-linearity of the time-reversal operator changes the sign of the complex interaction term in the Lagrangian. Hence, the Lagrangian is $\eta_P^{-1}\eta_T$-symmetric:
\be\label{eq_etaP_etaT_symmetric}
\eta_P^{-1}\eta_T:\:\lag(x)\longmapsto \lag^{\eta_P^{-1} \eta_T}(x_{PT})=\lag (x)\:,\:\:\:(\heta_P^{-1}\heta_T)\hlag (x_{PT})(\heta_P^{-1}\heta_T)^{-1}=\hlag (x)\:.
\ee
In subsection $3.6$, we will show that the operators $\heta_P$ and $\heta_T$ are related to the \comma standard' parity $\hpr$ and time-reversal $\htr$ transformations, such that $\heta_P^{-1}\heta_T=\hpr\htr$, meaning the Lagrangian is, in fact, still $PT$ symmetric with respect to the \comma standard' $PT$ transformation.

The pseudo-reality condition in Eq.~\eqref{eq_pseudo_reality_def} relates the field $\hat{\phi}$ to its Hermitian conjugate $\hat{\phi}^\dag$ via the linear operator $\heta_{P}$ through
\be\label{eq_psc_P}
\hph^\dag (x)=-\heta_{P} \hph (x_P)\heta_{P}^{-1}=-\eta_{P}[\hph (x)]\:,
\ee
and via the anti-linear operator $\heta_T$ through
\be\label{eq_psc_T}
\hph^\dag (x)=\heta_{T} \hph (x_T)\heta_{T}^{-1}=\eta_{T}[\hph (x)]\:.\ee
As we will show in the following subsections, both of these conditions are equivalent and give a single constraint on the field $\hat{\phi}$.


\subsection{Finding $\eta$}

To make explicit the constraints provided by the pseudo-reality conditions in Eqs.~\eqref{eq_psc_P} and \eqref{eq_psc_T}, we will now determine the explicit forms of the transformations $\eta_P$ and $
\eta_T$. As we will see, in the case where the non-Hermiticity is controlled by a small parameter, this can be achieved by obtaining a differential equation for the transformed fields that resembles the equation of motion for the field.

We begin with the \comma parity-like' transformation $\eta_P$, given in Eq.~\eqref{eq_correct_parity},
\be\label{eq_eta_P}
    \eta_P:\:\:\phi(x)\longmapsto \eta_P[\phi(x)]=-\tilde{\phi}(x)\:.
\ee
The field $\eta_P[\phi(x)]$ solves the same equation of motion~\eqref{eq_motion_dual} as the field $-\td{\phi}(x)$, i.e.,
\be\label{eq_eq_motion_eta}
\Box_{x}\eta_P[\phi(x)]+m^2\eta_P[\phi(x)]+\frac{ig}{2}\qty(\eta_P[\phi(x)])^2=0\:.
\ee
This equation is non-linear, but it can be solved perturbatively in orders of the coupling constant $g$. Expanding  $\eta_P=\eta_P^{(0)}+g\eta_P^{(1)}+g^2\eta_P^{(2)}$ at order $g^2$ gives us
\begin{subequations}
\begin{align}\label{eq_eta0}
    &\qty(\Box_x+m^2)\eta_P^{(0)}-\frac{ig}{2}\qty(\eta_P^{(0)})^2=0\:,\\
    \label{eq_eta1}&\qty(\Box_x+m^2)\eta_P^{(1)}+i\qty(\eta^{(0)})^2_P=0\:,\\
    \label{eq_eta2}&\qty(\Box_x+m^2)\eta_P^{(2)}+i\eta^{(0)}_P\eta^{(1)}_P=0\:,
\end{align}
\end{subequations}
where we have added and subtracted a term $ig\qty(\eta_P^{(0)})^2$ in the overall expansion, so that the equation for $\eta_P^{(0)}$ is dependent on $g$. We do this because we wish to find a functional $\eta_P[\phi(x)]$ which depends on the complex interacting field $\phi$ and not on the real free Klein-Gordon field $\phi_0$.

The equation~\eqref{eq_eta0} has a solution $\eta_P^{(0)}[\phi(x)]=-\phi(x)$. The remaining equations are solved using an Ansatz and, at order $g^2$, the map $\eta_P$ is found to be of the form
\begin{equation}\label{eq_etafP} \eta_P[\phi(x)]=-\phi (x)+ig \int {\rm d}^4y\:G_{0}(x,y)\phi^{2}(y)+g^2 \int {\rm d}^4y\:{\rm d}^4z\:G_{0}(x,y)\phi(y)G_{0}(y,z)\phi^2 (z)+\mathcal{O}(g^3)\:.
\end{equation}
Here, the two-point function $G_{0}(x,y)$ is the Green's function of the free Klein-Gordon equation~\eqref{eq_green}. The map $\eta_P$ bears some resemblance to the map derived in Refs.~\cite{Li:2023kpi, Li:2024xms}.\footnote{In Ref.~\cite{Li:2023kpi}, it is suggested that there exists an arbitrariness to the Hermitian theory that corresponds to the $i\phi^3$ theory in its $PT$-symmetric regime. We disagree with this conclusion, and it is readily confirmed that the apparent arbitrariness remaining in Eq.~(17) of Ref.~\cite{Li:2023kpi} is eliminated when performing the path integral. This is to say that the terms proportional to the unconstrained coefficients $c_2$, $c_4$ and $c_5$ that appear therein cancel at order $g^2$.}

The pseudo-reality condition~\eqref{eq_eta_P} gives the relationship between the field $\phi$ and its Hermitian conjugate $\phi^\dag$: 
\begin{equation}\label{eq_conj}
    \phi^{\dag}(x)=\phi (x)-ig \int {\rm d}^4y\:G_{0}(x,y)\phi^{2}(y)-g^2 \int {\rm d}^4y\:{\rm d}^4z\:G_{0}(x,y)\phi(y)G_{0}(y,z)\phi^2 (z)+\mathcal{O}(g^3)\;.
\end{equation}
This equation defines a hypersurface in the complex field space and reduces the number of degrees of freedom of the complex field $\phi$ to that of a real field, i.e., one.

The time-reversal-like map $\eta_T$ solves the equation of motion for $-\eta_P$. Hence, the solution is straightforward
\begin{equation}
    \eta_T[\phi(x)]=\phi (x)-ig \int {\rm d}^4y\:G_{0}(x,y)\phi^{2}(y)-g^2 \int {\rm d}^4y\:{\rm d}^4z\:G_{0}(x,y)\phi(y)G_{0}(y,z)\phi^2 (z)\:+\mathcal{O}(g^2)\;,
\end{equation}
and yields the same pseudo-reality condition  as Eq.~\eqref{eq_conj}.


\subsection{$\heta$ operator}
\label{sec:etaoperator}

Now that we know how the transformations $\eta_P$ and $\eta_T$ act on the field $\phi$, we wish to find the corresponding Hermitian operators $\heta_P$ and $\heta_T$, with
\be
\eta_{P}[\hph(x)]=\heta_P \hph(x_P)\heta^{-1}_P\qquad\text{and}\qquad \eta_{T}[\hph (x)]=\heta_T \hph(x_T)\heta^{-1}_T\:.
\ee

First, we need to choose a Green's function $G_0(x,y)$ in the expansion~\eqref{eq_etafP}. The equal-time commutation relations
\begin{subequations}
\begin{align}
&[\hph(t,\vx),\hp(t,\vy)]=i\delta^{(3)}(\vx-\vy)\:,\\
\label{eq:phphicomm}
&[\hph(t,\vx),\hph(t,\vy)]=[\hp(t,\vx),\hp(t,\vy)]=0\:,
\end{align}
\end{subequations}
are only preserved under the transformation $\eta$ for a causal and Hermitian Green's function. Here, $\hat{\pi}(x)=\frac{\partial\hlag(x)}{\partial\dot{\hat{\phi}}(x) }=\dot{\hat{\phi}}(x)$ is the momentum operator conjugate to $\hat{\phi}$. We recall from Sec.~\ref{eq:issues} that the field operator $\hph(x)$ and its Hermitian conjugate $\hph^\dag(x)$ do not commute at equal times, i.e.,
\be
\label{eq:phiphidagcomm}
[\hph(t,\vx),\hph^\dag (t,\vy)]\neq 0\:,
\ee
since $\hat{\phi}^\dag$ is not the field operator dual to $\hat{\phi}$. Indeed, it is the dual field operator $\hat{\tilde{\phi}}^\dag$ that commutes with $\hat{\phi}$ at equal times:
\be
[\hph (t,\vx),\thph{\dag}{}{t,\vy}]=0\:,
\ee
which is trivially consistent with Eq.~\eqref{eq:phphicomm} by virtue of the pseudo-reality condition~\eqref{eq_pseudo_reality_def}.

The causal Green's functions solving~\eqref{eq_green} are the retarded and the advanced propagators:
\begin{subequations}
\begin{align}\label{eq_R_A_prop}
G_R (x,y)=\left\{\begin{array}{l}
D(x,y)-D(y,x)\text{\:\:\:for\:\:\:}x^0>y^0 \\
0 \text{\:\:\:for\:\:\:} y^0>x^0
\end{array}\right.\:,\\
G_A (x,y)=\left\{\begin{array}{l}
0\text{\:\:\:for\:\:\:}x^0>y^0 \\
D(y,x)-D(x,y) \text{\:\:\:for\:\:\:} y^0>x^0
\end{array}\right.\:,
\end{align}
\end{subequations}
where $D(x,y)$ is the free Wightman function
\be
D(x,y)=\int\! \frac{{\rm d}^3 \vec{p}}{(2\pi)^3}\: \frac{1}{2iE_{\vec{p}}}\,e^{-ip\cdot(x-y)}
\:.
\ee
In the remainder of this section, we choose to work with the retarded propagator. However, we could likewise use the advanced propagator.

We will show below that the operator $\heta_P$ can be written as a combination of an operator $\heta_g$, which can be obtained as a perturbative expansion in the coupling $g$, and the \comma standard' parity operator $\hpr$. The \comma standard' parity operation has the usual form
\be
\hpr (t)\hph(t,\vx)\hpr^{-1}(t)=-\hph (t,-\vx)\:,
\ee
with
\be
\hpr(t)=e^{\frac{\pi}{2}\int {\rm d}^3\vec{z} (\hph(t,\vec{z})\hp(t,-\vec{z})+\hph(t,\vec{z})\hp(t,\vec{z})  )}\:.
\ee
The operator $\heta_g$ is less trivial, and the full calculation is provided in Appendix~\ref{sec:etacalc}. We are, however, looking for an expansion of the form
\begin{align}\label{eq_req_expansion}
\heta_g(t,t_0)\hph(t,\vx)\heta^{-1}_g(t,t_0)&=\hph(t,\vx)-ig\int^{t}_{t_0}{\rm d}y^0 \int {\rm d}^3\vec{y}\:G_{R}(x,y)\hph^2(y)\nonumber\\
&-g^2\int^{t}_{t_0}{\rm d}y^0\int^{y^0}_{t_0}{\rm d}z^0\int {\rm d}^3\vec{y}\:{\rm d}^3\vec{z}\:G_{R}(x,y)\hph(y)G_{R}(y,z)\hph^2(z)\nonumber\\
&+i g^2 
\int^{t}_{t_0}{\rm d}y^0\int^{y^0}_{t_0}{\rm d}z^0\int {\rm d}^3\vec{y}\:{\rm d}^3\vec{z}\: G_{R}(x,y)G_{R}(y,z)G_{R}(y,z)\hph (z)\:.
\end{align}
Combining this with the \comma standard' parity operator, gives us the required expansion in Eq.~\eqref{eq_etafP}:
\begin{align}
    \heta_g(t,t_0)\hpr(t)\hph(t,-\vx)&\hpr^{-1}(t)\heta_g^{-1}(t,t_0)=-\heta_g(t,t_0)\hph(t,\vx)\heta_g^{-1}(t,t_0)\nonumber\\
    &=-\hph(t,\vx)+ig\int^{t}_{t_0}{\rm d}y^0 \int {\rm d}^3\vec{y}\:G_{R}(x,y)\hph^2(y)\nonumber\\
&+g^2\int^{t}_{t_0}{\rm d}y^0\int^{y^0}_{t_0}{\rm d}z^0\int {\rm d}^3\vec{y}\:{\rm d}^3\vec{z}\:G_{R}(x,y)\hph(y)G_{R}(y,z)\hph^2(z)\nonumber\\
&-i g^2 
\int^{t}_{t_0}{\rm d}y^0\int^{y^0}_{t_0}{\rm d}z^0\int {\rm d}^3\vec{y}\:{\rm d}^3\vec{z}\: G_{R}(x,y)G_{R}(y,z)G_{R}(y,z)\hph (z)\:.
\end{align}
As we will now show, the last term is a loop correction of order $\hbar$ and thus vanishes in the semi-classical limit,  $\hbar\mapsto 0$, relevant to Eq.~\eqref{eq_etafP}.

Taking the expectation value of Eq.~\eqref{eq_req_expansion} and using Wick's theorem, we obtain
\begin{align}
\label{eq_expectation}
-\big<\heta_g(t,t_0)\hph(t,\vx)&\heta_g^{-1}(t,t_0)\big>=-\phi(t,\vx)-ig\int^{t}_{t_0}{\rm d}y^0 \int {\rm d}^3\vec{y}\:G_{R}(x,y) [\phi^2(y)+\hbar G_F(y,y)]\nonumber\\
&-g^2\int^{t}_{t_0}{\rm d}y^0\int^{y^0}_{t_0}{\rm d}z^0\int {\rm d}^3\vec{y}\:{\rm d}^3\vec{z}\:G_{R}(x,y)G_{R}(y,z)\nonumber\\&\times[\phi(y)\phi^2(z)+\hbar\phi(y)G_F(z,z)+2\hbar G_F(y,z)\phi(z)]\nonumber\\
&+i \hbar g^2 
\int^{t}_{t_0}{\rm d}y^0\int^{y^0}_{t_0}{\rm d}z^0\int {\rm d}^3\vec{y}\:{\rm d}^3\vec{z}\: G_{R}(x,y)G_{R}(y,z)G_{R}(y,z)\phi (z)\:,
\end{align}
where we have reinstated factors of $\hbar$ for clarity and
\begin{align}
G_F(x,y)&=\big<\mathrm{T}[\hat{\phi}(x)\hat{\phi}(y)]\big>\nonumber\\&=i\theta(x^0-y^0)D(x,y)+i\theta(y^0-x^0)D(y,x)+\mathcal{O}(g)
\nonumber\\&=\int\frac{{\rm d}^4 p}{(2\pi)^4}\frac{i e^{-ip\cdot(x-y)}}{p^2-m^2+i\epsilon}+\mathcal{O}(g)
\end{align}
is the Feynman propagator. (Note that the Feynman propagator $G_F$ is the full, dressed Feynman propagator, whereas the retarded propagator $G_R$ is tree level.) It is now clear that the terms of order $\hbar$ in Eq.~\eqref{eq_expectation} can be identified as loop corrections. As we take the classical limit $\hbar \mapsto 0$, these terms vanish, and we recover the classical expansion in Eq.~\eqref{eq_etafP} after identifying the $c$-number field $\phi(x)$ with $\big<\hat{\phi}(x)\big>$ in the same limit. Notice that the combination $iG_R(y,z)[2G_F(y,z)-iG_R(y,z)]=G_F^2(y,z)+D^2(y,z)$, obtained from terms in the second and third lines, is nothing other than the retarded bubble diagram, as we would expect. The non-commutativity of the field and its Hermitian conjugate is now readily confirmed, and, at order $g$, we have
 \begin{equation}
     \left[\hat{\phi}(t,\vec{x}),\hat{\phi}^{\dag}(t,\vec{y})\right]=-ig\int{\rm d}^4z\;G_R(x,z)G_R(y,z)\hat{\phi}(z)\;,
 \end{equation}
where $x^0=y^0=t$.

Given that the perturbation series for the solution for the transformation [Eq.~\eqref{eq_etafP}] is similar to that of the interacting Klein-Gordon equation, we find (see Appendix~\ref{sec:etacalc}) that the series expansion is generated by an operator similar to the time-evolution operator; specifically,
\be\label{eq_hetag}
\heta_g(t,t_0)=\hu^{-1}\qty(-\frac{1}{3},t,t_0)\:,
\ee
where we define the operator $\hu$ via
\begin{align}
\hu\qty(\al,t,t_0)\df T\qty[e^{\al g \int^t_{t_0}{\rm d}y^0\int {\rm d}^3\vec{y}\:\hph^3(y^0,\vy)  }]\:,
\end{align}
and
\begin{align}
\hu^{-1}\qty(\al,t,t_0)=\bar{T}\qty[e^{-\al g\int^t_{t_0}{\rm d}y^0 \int {\rm d}^3\vec{y}\:\hph^3(y^0,\vy)  }]\:,
\end{align}
for some constant $\al\in\cn$. Here, $T$ is the time-ordering operator and $\bar{T}$ is the anti-time-ordering operator. Note that $\hu(\frac{1}{3!},t,t_0)$ is (when replacing $\hat{\phi}$ by $\hat{\phi}_0$) the usual time-evolution operator whose action on the free Klein-Gordon field $\phi_0$ gives the series expansion in Eq.~\eqref{eq_phi_expansion}.

In Appendix~\ref{sec:etaHerm}, we show that, at order $g^2$, 
the operators $\heta_g$ and $\heta_P$ are Hermitian:
\be
\heta_g^\dag (t,t_0)=\heta_g(t,t_0)\text{\:\:\:and\:\:\:}  \heta^\dag_P(t,t_0)=\heta_P(t,t_0)\:,
\ee
and the \comma standard' parity operator is $\heta_P$-pseudo-Hermitian, i.e.,
\be
\hpr^\dag (t)=\heta_P (t,t_0)\hpr (t)\heta_P^{-1}(t,t_0)\;.
\ee
Hence, the \comma parity-like' operator that makes our Lagrangian pseudo-Hermitian is given by
\be
\heta_P(t,t_0)=\heta_g(t,t_0)\:\hpr(t)\:,
\ee
as claimed above.

In a similar manner, we can write the \comma time-reversal-like' operator $\heta_T=\heta_g \htr$ as a combination of the \comma standard' time-reversal operator
\be
\htr(t) \hph (t,\vx)\htr^{-1}(t)=\hph (-t,\vx)
\ee
and the operator $\heta_g$ in Eq.~\eqref{eq_hetag}. Thus, the Lagrangian is anti-pseudo-Hermitian with respect to the anti-linear operator $\heta_T$:
\be
\heta_T(t,t_0)=\heta_g (t,t_0)\htr (t).
\ee
Given that the Lagrangian is $\eta_P^{-1}\eta_T$-symmetric [see Eq.~\eqref{eq_etaP_etaT_symmetric}], we see that it is also $PT$ symmetric with respect to the \comma standard' $PT$ transformation, since $\heta_P^{-1}\heta_T=\hpr\htr$.

Thus, we have demonstrated that by adopting a first-principles approach and introducing the pseudo-reality condition, we can achieve a self-consistent formulation of the \(i\phi^3\) theory. This framework resolves the contradictions arising from Euler-Lagrange equations and \comma standard' parity and time-reversal transformations and ensures that the theory remains well defined with a real spectrum under exact \(PT\) symmetry.


\section{Generalisation to higher spins}

In the preceding sections, we have focussed on the simplest, spin-zero representation. We now turn our attention to higher-spin representations. Of particular interest is the coupling to gauge fields, and we begin from a well-studied theory composed of two complex scalar fields with an anti-Hermitian mass mixing~\cite{Alexandre:2017foi, Alexandre:2020gah, Sablevice:2023odu}. Doing so will provide an illustration of how pseudo-reality allows us to gauge non-Hermitian quantum field theories and to couple them to gravity consistently. Moreover, this two complex scalar field theory falls under the second approach to constructing non-Hermitian quantum field theories, constructed ab-initio from non-Hermitian composite operators. We will see, however, that pseudo-reality also allows to interpret this theory as the analytic continuation of a Hermitian theory.


\subsection{Two complex scalar field theory}

We start with the following free field theory (see, e.g., Refs.~\cite{Alexandre:2017foi, Alexandre:2020gah, Sablevice:2023odu}):
\begin{equation}
\mathcal{L}=\partial_{\mu}\tilde{\Phi}^{\dag}\partial^{\mu}\Phi-\tilde{\Phi}^{\dag}M^2\Phi\;,
\end{equation}
where the field $\Phi\equiv (\phi_1, \phi_2)$ is composed of two complex scalar fields $\phi_1$ and $\phi_2$. As the original Lagrangian, composed of $\Phi$ and $\Phi^\dag$ is $PT$ symmetric and parity-pseudo-Hermitian, the dual field $\tilde{\Phi}^{\dag}\equiv(\tilde{\phi}_1^{\dag},\tilde{\phi}_2^{\dag})$ can be given via the parity transformation $\hpr$
\begin{equation}
    \hat{\tilde{\Phi}}^{\dag}(x)=\hat{\mathcal{P}}^{-1}\hat{\Phi}^{\dag}(x_P)\hat{\mathcal{P}}\:\Pi\;,\qquad \Pi=\begin{pmatrix} +1 & 0 \\ 0 & -1\end{pmatrix}\;.
\end{equation}
Here, $\Pi$ is the parity matrix, reflecting the scalar and pseudo-scalar nature of the fields $\phi_1$ and $\phi_2$ respectively. The mass matrix has skew-symmetric off-diagonals and is $\Pi$-pseudo-Hermitian, with
\begin{equation}
    M^2=\begin{pmatrix} m_1^2 & m_5^2 \\ -m_5^2 & m_2^2\end{pmatrix}\;,\qquad M^{2^\dag}=\Pi M^2 \Pi^{-1}\:,
\end{equation}
where $m_1^2$, $m_2^2$ and $m_5^2$ real and positive.

We will now show that the field $\hat{\Phi}$ is a complex field that can be decomposed into two pseudo-real fields. We start by writing
\begin{equation}
    \hat{\phi}_j=\frac{1}{\sqrt{2}}\Big(\hat{\chi}_{j1}+i\hat{\chi}_{j2}\Big)\;,\qquad \text{and}\qquad \hat{\tilde{\phi}}^{\dag}_j=\frac{1}{\sqrt{2}}\Big(\hat{\tilde{\chi}}^{\dag}_{j1}-i\hat{\tilde{\chi}}^{\dag}_{j2}\Big)\;.
\end{equation}
Here, we take the fields $\hat{\chi}_{jk}\in \mathbb{C}$, following an approach similar to Ref.~\cite{Mannheim:2018dur} (cf.~the alternative approach of Ref.~\cite{Fring:2019hue}, where the field is decomposed into two real components). If the field components $\hat{\chi}_{jk}$ and $\hat{\tilde{\chi}}_{jk}^{\dag}$ are unrelated to each other, then we have eight degrees of freedom, rather than the four that we would expect for two complex scalar fields. We recover four degrees of freedom by imposing pseudo-reality on the field components $\hat{\chi}_{jk}$, namely that
\begin{equation}
    \hat{\tilde{\chi}}^{\dag}_{jk}(x)=\hat{\chi}_{jk}(x)\;.
\end{equation}
The Lagrangian then takes the form
\begin{align}
    \hat{\mathcal{L}}=\frac{1}{2}\sum_{j,k}\partial_{\mu}\hat{\chi}_{jk}\partial^{\mu}\hat{\chi}_{jk}-\frac{1}{2}m_1^2\left(\hat{\chi}_{11}^2+\hat{\chi}_{12}^2\right)-\frac{1}{2}m_2^2\left(\hat{\chi}_{22}^2+\hat{\chi}_{21}^2\right)-im_5^2\left(\hat{\chi}_{11}\hat{\chi}_{22}-\hat{\chi}_{12}\hat{\chi}_{21}\right)\;,
\end{align}
and the resulting equations of motion for the fields $\hat{\chi}_{jk}$ are
\begin{subequations}
    \begin{align}
    \Box\hat{\chi}_{11}+m_1^2\hat{\chi}_{11}+im_5^2\hat{\chi}_{22}=0\;,\\
    \Box\hat{\chi}_{12}+m_1^2\hat{\chi}_{12}-im_5^2\hat{\chi}_{21}=0\;,\\
    \Box\hat{\chi}_{21}+m_2^2\hat{\chi}_{21}-im_5^2\hat{\chi}_{12}=0\;,\\
    \Box\hat{\chi}_{22}+m_2^2\hat{\chi}_{22}+im_5^2\hat{\chi}_{11}=0\;.
    \end{align}
\end{subequations}
This theory is analytically continued from a Hermitian theory by taking $m_5^2\to im_5^2$, and we see immediately, as for the $i\phi^3$ theory discussed in Sec.~\ref{sec:iphi3_theory}, that the fields $\hat{\chi}_{jk}$ cannot be real. We now need only to ensure the self-consistency of the pseudo-reality of the fields $\hat{\chi}_{jk}$.

The pseudo-reality conditions for the components $\hat{\chi}_{jk}$ are given via the parity transformation:
\begin{subequations}
\begin{align}
\hat{\td{\chi}}^{\dag}_{11}&=\hat{\mathcal{P}}^{-1}\hat{\chi}^\dag_{11}\hat{\mathcal{P}}=\hat{\chi}_{11}\;,\\
\hat{\td{\chi}}^{\dag}_{12}&=\hat{\mathcal{P}}^{-1}\hat{\chi}^\dag_{12}\hat{\mathcal{P}}=\hat{\chi}_{12}\;,\\
\hat{\td{\chi}}^{\dag}_{21}&=-\hat{\mathcal{P}}^{-1}\hat{\chi}^\dag_{21}\hat{\mathcal{P}}=\hat{\chi}_{21} \;,\\
\hat{\td{\chi}}^{\dag}_{22}&=-\hat{\mathcal{P}}^{-1}\hat{\chi}^\dag_{22}\hat{\mathcal{P}}=\hat{\chi}_{22}\;.
\end{align}
\end{subequations}
Note that, unlike the case of the $i\phi^3$ theory, where the degree of non-Hermiticity is controlled by the small coupling constant $g$, we cannot obtain an expression $\eta[\hat{\chi}_{jk}]$ in terms of the original components $\hat{\chi}_{jk}$.


\subsection{Gauged model}

A number of attempts have been made in the literature to gauge similar scalar models that incorporate two complex scalar fields~\cite{Alexandre:2018xyy, Millington:2019dfn}. There are two possibilities: (i) couple the gauge field to a Hermitian but non-conserved current or (ii) couple the gauge field to a conserved but non-Hermitian current. Method (i) leads to an inconsistent Maxwell equation and breaks the gauge invariance of the Lagrangian. Method (ii) is in contradiction with the reality of the gauge field. We will show in this section  that method (ii) is consistent so long as the gauge field is taken to be a pseudo-real vector field.

To see this, we minimally couple a U(1) gauge field to the doublet $\Phi = (\phi_1, \phi_2)$ of two complex scalar fields as follows:
\begin{equation}
    \mathcal{L}=-\frac{1}{4}F_{\mu\nu}\tilde{F}^{\dag\mu\nu}-\frac{1}{2\xi}(\partial_{\mu}A^{\mu})(\partial_{\mu}\tilde{A}^{\dag\mu})+\tilde{D}_{\mu}^{\dag}\tilde{\Phi}^{\dag}D^{\mu}\Phi-\tilde{\Phi}^{\dag}M^2\Phi\;,
\end{equation}
where $F_{\mu\nu}=\partial_{\mu}A_{\nu}-\partial_{\nu}A_{\mu}$ is the usual field strength tensor, and the gauge covariant derivatives are
\begin{equation}
    D_{\mu}=\partial_{\mu}-ieA_{\mu}\;,\qquad \tilde{D}_{\mu}^{\dag}=\partial_{\mu}+ie\tilde{A}_{\mu}^{\dag}\;,
\end{equation}

Let us now consider the operator-valued conserved current
\begin{equation}\hat{j}^{\nu}=ie\left[\hat{\tilde{\Phi}}^{\dag}\overset{\rightarrow}{D^{\nu}}\hat{\Phi}-\hat{\tilde{\Phi}}^{\dag}\overset{\leftarrow}{\tilde{D}^{\nu\dag}}\hat{\Phi}\right]\;.
\end{equation}
This current can be written in terms of the pseudo-real components as follows:
\begin{align}
    \hat{j}^{\nu}&=-e\left[\hat{\chi}_{11}\overset{\rightarrow}{\partial^{\nu}}\hat{\chi}_{12}-\hat{\chi}_{11}\overset{\leftarrow}{\partial^{\nu}}\hat{\chi}_{12}+\hat{\chi}_{21}\overset{\rightarrow}{\partial^{\nu}}\hat{\chi}_{22}-\hat{\chi}_{21}\overset{\leftarrow}{\partial^{\nu}}\hat{\chi}_{22}\right]\nonumber\\&\phantom{=}+\frac{1}{2}e^2\left(\hat{A}^{\nu}+\hat{\tilde{A}}^{\nu\dag}\right)\left[\hat{\chi}_{11}^2+\hat{\chi}_{12}^2+\hat{\chi}_{21}^2+\hat{\chi}_{22}^2\right]\;.
\end{align}
This current is pseudo-real if the gauge field satisfies the pseudo-reality condition
\begin{equation}
    \hat{\tilde{A}}^{\mu\dag}(x)=\hat{\mathcal{P}}^{-1}\hat{A}^{\nu\dag}(x_P)\hat{\mathcal{P}}\Pi_{\nu}^{\phantom{\nu}\mu}=\hat{A}^{\mu}(x)\;,
    \label{eq_A_pseudo_real}
\end{equation}
where $\Pi_{\nu}^{\phantom{\nu}\mu}=\delta_{\nu}^{\phantom{\nu}\mu}$ in this case. 

The following operator-valued Maxwell equations are now all internally consistent:
\begin{subequations}
\begin{align}
    \Box \hat{A}^{\nu}-\left(1-\frac{1}{\xi}\right)\partial^{\nu}\partial_{\mu}\hat{A}^{\mu}=\hat{j}^{\nu}\;,\\
    \Box \hat{\tilde{A}}^{\nu}-\left(1-\frac{1}{\xi}\right)\partial^{\nu}\partial_{\mu}\hat{\tilde{A}}^{\mu}=\hat{\tilde{j}}^{\nu}\;,\\
    \Box \hat{A}^{\nu\dag}-\left(1-\frac{1}{\xi}\right)\partial^{\nu}\partial_{\mu}\hat{A}^{\mu\dag}=\hat{j}^{\nu\dag}\,,
\end{align}
\end{subequations}
and the pseudo-reality condition~\eqref{eq_A_pseudo_real} implies that the photon field $A_\mu$, although complex, possesses only two propagating degrees of freedom.

We anticipate that this consistent pseudo-Hermitian formulation of a U(1) gauge theory will allow to revisit existing discrepancies in the treatments of the Higgs mechanism in similar non-Hermitian field theories (cf.~Refs.~\cite{Mannheim:2018dur, Alexandre:2018xyy, Alexandre:2019jdb}). This conjecture may be checked in future work. 


\subsection{Coupling to gravity}

Having successfully gauged the two complex scalar model in the previous section, we speculate on how to couple non-Hermitian field theories to spin-2 fields. Consider, e.g., the free scalar sector described above. We can couple this theory minimally to Einstein gravity through the matter action
\begin{align}
    S=\int{\rm d}^4x\sqrt{-g}\left[g^{\mu\nu}\partial_{\mu}\tilde{\Phi}^{\dag}\partial_{\nu}\Phi-\tilde{\Phi}^{\dag}M^2\Phi\right]\;,
\end{align}
where $g_{\mu\nu}$ is the metric tensor. The resulting energy-momentum tensor is
\begin{equation}
    T^{\mu\nu}=\frac{1}{2}\partial^{(\mu}\tilde{\Phi}^{\dag}\partial^{\nu)}\Phi-g^{\mu\nu}\mathcal{L}\;,
\end{equation}
where the parenthesis indicates symmetrisation of the Lorentz indices.
This energy-momentum tensor is not Hermitian, but it is pseudo-Hermitian with respect to the transformation $\eta$, with $\eta=\mathcal{P}$ in this case. The Einstein equation
\begin{equation}
    G_{\mu\nu}=R_{\mu\nu}-\frac{1}{2}g_{\mu\nu}R=\kappa T_{\mu\nu}
\end{equation}
then implies that the Einstein tensor $G_{\mu\nu}$ is not Hermitian in general. Nevertheless, the Einstein equation is consistent as long as the metric $g_{\mu\nu}$, Ricci scalar $R$ and Ricci tensor $R_{\mu\nu}$ are pseudo-real with respect to the same transformation $\eta$, i.e.,
\begin{equation}
    \tilde{R}^{\dag}_{\mu\nu}-\frac{1}{2}\tilde{g}_{\mu\nu}^{\dag}\tilde{R}^{\dag}=R_{\mu\nu}-\frac{1}{2}g_{\mu\nu}R\;,
    \label{eq_R_pseudo}
\end{equation}
with the metric satisfying
\begin{equation}\tilde{g}^{\dag}_{\mu\nu}(x)=g_{\mu\nu}(x)\;,
    \label{eq_g_pseudo}
\end{equation}
where, at the operator level, we would have $\hat{\tilde{g}}^{\dag}_{\mu\nu}(x)=\hat{\eta}^{-1}\hat{g}^{\dag}_{\alpha\beta}(x_{\eta})\hat{\eta}\Pi^{\alpha\beta}_{\mu\nu}$, with $\Pi^{\alpha\beta}_{\mu\nu}$ fixed by the pseudo-Hermiticity of the Lagrangian.

Most notably, this construction requires the metric to be complex. While the idea of complex metrics is not new (see, e.g., Ref.~\cite{Witten:2021nzp} and references therein), a comprehensive analysis of the implications of pseudo-reality for complex metrics is beyond the scope of this work. Interestingly, however, the pseudo-reality of the curvature tensors~\eqref{eq_R_pseudo} and the metric~\eqref{eq_g_pseudo} is determined by the pseudo-Hermiticity properties of all non-Hermitian sectors to which they couple, as was the case for the vector field in the previous section. We leave further study of these pseudo-real geometries to future work, but we anticipate that the observations in this work pave the way to consistent formulations of non-Hermitian cosmological models.


\section{Conclusions}

We have presented a general framework within which Hermiticity puzzles in pseudo-Hermitian quantum field theories can be resolved. We refer to this property as \emph{pseudo-reality}. We first introduced the concept of pseudo-reality for complex numbers and then generalized it to quantum fields, promoting a real field defined on the real axis to a pseudo-real field defined on a contour in the complex plane. This approach resolves a long-standing inconsistency in the minimal coupling of pseudo-Hermitian sectors to gauge fields. We have focused here on a U(1) gauge theory and indicated how pseudo-reality may enable a consistent coupling to gravity.  Interestingly, the pseudo-reality conditions depend on the map $\eta$ that determines the pseudo-Hermiticity of the full Hamiltonian. Thus, the pseudo-reality condition of a given bosonic field is determined by the non-Hermitian sectors to which it couples, making clear the subtleties of treating interacting pseudo-Hermitian quantum field theories. We leave in-depth study of such theories and their couplings to gravity to future work.


\acknowledgments

The authors thank the organisers and participants of the meeting ``Applications of Field Theory to Hermitian and Non-Hermitian Systems'', hosted by King's College London, 10--13 September 2024, where this work was first presented. P.M. would like to thank Jean Alexandre, John Ellis, Andreas Fring, Philip Mannheim, Dries Seynaeve, Takano Taira and Andreas Trautner for helpful discussions. This work was supported by the International Exchanges project of the Royal Society No.~IES\textbackslash R3\textbackslash203069 and by the International Emerging Actions (IEA) project ``Non-Hermitian field theories for particle and solid-state physics / TH\'ECHNO'' of CNRS. The work of P.M.~was supported by the Science and Technology Facilities Council (STFC) [Grant No.~ST/X00077X/1] and a UKRI Future Leaders Fellowship [Grant No.~MR/V021974/2]. The work of E.S. was supported by the University of Manchester.


\section*{Data Access Statement}

No data were created or analysed in this study.

\appendix

\renewcommand\theequation{\thesection.\arabic{equation}}


\section{Calculating the $\heta$ operator}
\label{sec:etacalc}

We note that the expansion of the $\eta_P$ transformation in Eq.~\eqref{eq_etafP} has the same terms as the expansion of $\phi$ in Eq.~\eqref{eq_phi_expansion}. Hence, we make an Ansatz that the operator $\heta_P$ should be of the form of the time-evolution operator, but with differing coefficients and expressed in terms of the full, interacting field $\hat{\phi}$. We define an operator $\hu$
\begin{align}
\hu\qty(\al,t,t_0)\df T\qty[e^{\al g \int^t_{t_0}{\rm d}y^0\int {\rm d}^3\vec{y}\:\hph^3(y^0,\vy)  }]\:,
\end{align}
and
\begin{align}
\hu^{-1}\qty(\al,t,t_0)=\bar{T}\qty[e^{-\al g\int^t_{t_0}{\rm d}y^0 \int {\rm d}^3\vec{y}\:\hph^3(y^0,\vy)  }]\:,
\end{align}
for some constant $\al\in\cn$. Here, $T$ is the time-ordering operator and $\bar{T}$ is the anti-time-ordering operator. 

We proceed by expanding
\be\label{eq_expansion}
\hu^{-1}(\al,t,t_0)\hph(t,\vx)\hu (\al,t,t_0)=\hph_x-\al g\intax{y}\qty[\hph^3_y,\hph_x]+\al^2g^2\intax{y}\intay{z}\qty[\hph^3_z,\qty[\hph^3_y,\hph_x]]\;,
\ee
where we define shorthand notations for the integral $\intax{y}\df\int_{t_0}^{x^0}{\rm d}y^0\:\int {\rm d}^3\vec{y}$ and the field $\hph_x \df \hph(x^0,\vec{x})$.

We note that the commutator $[\hph_x,\hph_y]$ corresponds to that of the \emph{interacting} field $\hph$ and not to that of the free Klein-Gordon field $\hph_0$. However, we can expand the interacting field as a series in the coupling constant $g$. For our purposes, we only need to perform this expansion at first order in $g$:
\be
\hph_x=\hph_{0x}+\frac{ig}{2}\intax{y}G_{Rxy}\hph_{0y}^2+\mathcal{O}(g^2)\;.
\ee
This relation can be inverted (again, at linear order in $g$) to obtain
\be
\hph_{0x}=\hph_x-\frac{ig}{2}\intax{y}G_{Rxy}\hph_y^2+\mathcal{O}(g^2)\;.
\ee
Herein, $G_{Rxy}\df G_{R}(x,y)$ denotes the retarded propagator. We also denote the Pauli--Jordan function by $G_{xy}\df D_{xy}-D_{yx}$, which only becomes the retarded propagator with appropriate integral limits.

We calculate the commutators for the expansion in Eq.~\eqref{eq_expansion} at order $g$ to obtain
\begin{align}
[\hph_y,\hph_x]&=\qty[\hph_{0y}+\frac{ig}{2}\intay{z}G_{Ryz}\hph_{0z}^2+\mathcal{O}(g^2),\hph_{0x}+\frac{ig}{2}\intax{z}G_{Rxz}\hph_{0z}^2+\mathcal{O}(g^2)]\nonumber\\
&=\qty[\hph_{0y},\hph_{0x}]-g\intax{z}G_{Rxz}G_{yz}\hph_{0z}+g\intay{z}G_{xz}G_{Ryz}\hph_{0z}+\mathcal{O}(g^2)\;,
\end{align}
and invert the expression for $\phi$ to obtain
\be
[\hph_y,\hph_x]=-iG_{xy}-g\intax{z}G_{Rxz}G_{yz}\hph_{z}+g\intay{z}G_{xz}G_{Ryz}\hph_{z}\;.
\ee
Hereafter, we omit the terms that are subleading in the coupling constant $g$.

The last relation allows us to calculate the commutator:
\begin{align}
[\hph^3_y,\hph_x]&=-3iG_{xy}\hph^2_y-3g\intax{z}G_{Rxz}G_{yz}\hph_z\hph^2_y+3g\intay{z}G_{xz}G_{Ryz}\hph_z\hph^2_y\nonumber\\
&\phantom{=}-3ig\intax{z}G_{Rxz}G_{yz}G_{yz}\hph_y+3ig\intay{z} G_{xz}G_{Ryz}G_{yz}\hph_y\;.
\end{align} 
For the third term in Eq.~\eqref{eq_expansion}, we only need the commutator at $0$th order in g:
\be
[\hph^3_z,[\hph^3_y,\hph_x]]=-18G_{xy}G_{yz}\hph_y\hph^2_z+18iG_{xy}G_{yz}G_{yz}\hph_z\;.
\ee

Plugging the commutators into Eq.~\eqref{eq_expansion}, we get
\begin{align}
\hu^{-1}(\al,t,t_0)\hph(t,\vx)\hu (\al,t,t_0)&=\hph_x+3i \al g\intax{y}G_{Rxy}\hph^2_y\nonumber\\
&\phantom{=}-3g^2\al (1+6\al)\intax{y}\intay{z}G_{Rxy}G_{Ryz}\hph_y\hph^2_z\nonumber\\&\phantom{=}+3i\al g^2(1+6\al)\intax{y}\intay{z}G_{xy}G_{yz}G_{yz}\hph_z\;.
\end{align}
Finally, taking $\al=-\frac{1}{3}$, we get the operator $\heta_g(t)=\hu^{-1}\qty(-\frac{1}{3},t,t_0)$ in Eq.~\eqref{eq_req_expansion}, giving
\begin{align}
\heta_g(t,t_0)\hph(t,\vx)\heta_g^{-1} (t,t_0)&=\hph_x-i g\intax{y}G_{Rxy}\hph^2_y\nonumber\\
&\phantom{=}-g^2\intax{y}\intay{z}G_{Rxy}G_{Ryz}\hph_y\hph^2_z\nonumber\\
&\phantom{=}+i g^2\intax{y}\intay{z}G_{xy}G_{yz}G_{yz}\hph_z\;,
\end{align}
with the last term appearing due to quantum loop corrections. It is proportional to $\hbar$ and vanishes in the classical limit $\hbar \mapsto 0$ (see Section~\ref{sec:etaoperator}).


\section{Proving the Hermiticity of $\heta$}
\label{sec:etaHerm}

In this Appendix, we show that $\heta_P$ and $\heta_g$ are Hermitian while $\hpr$ is $\heta_P$-pseudo-Hermitian.

We expand the operator $\hu$ at order $g^2$:
\begin{align}
    \hu(\al,t,t_0)=T\qty[e^{\al\intax{y}\hph^3_y}]=1+\al g\intax{y}\hph^3_y+\al^2 g^2 \intax{y}\intay{z}\hph^3_y\hph^3_z\;.
\end{align}
Taking its Hermitian conjugate, we have
\begin{align}
    \label{eq_U_dag}
    \hu^{\dag}(\al,t,t_0)=1+\al g\intax{y}\hph^{3^\dag}_y+\al^2 g^2 \intax{y}\intay{z}\hph^{3^\dag}_z\hph^{3^\dag}_y\;.
\end{align}
The pseudo-reality condition gives us an expansion of $\hph^\dag$ in terms of $\hph$ [see Eq.~\eqref{eq_conj}]. At order $g$, we have
\begin{align}
    \hph^\dag_x=\hph_x-ig\intax{y}G_{Rxy}\hph^2_y\;.
\end{align}
Hence, after commuting all $\hph_x$ to the left of $\hph_y$, the interaction term is
\begin{align}
    \hph^{3^\dag}_x=\hph^3_x-3ig\intax{y}G_{Rxy}\hph^2_x\hph^2_y-6g\intax{y}G_{Rxy}G_{xy}\hph_x\hph_y+2ig\intax{y}G_{Rxy}G_{xy}G_{xy}\;.
\end{align}
We insert this expression into Eq.~\eqref{eq_U_dag} and switch the order of $\hph^3_z\hph^3_y$ in the last term using the commutator
\begin{align}
    [\hph^3_z,\hph^3_y]=-9iG_{yz}\hph^2_y\hph^2_z-18G_{yz}G_{yz}\hph_y\hph_z+6iG_{yz}G_{yz}G_{yz}\;.
\end{align}
Thus, we get $\hu^\dag$ at order $g^2$
\begin{align}
    \hu^\dag (\al,t,t_0)&=1+\al g\intax{y}\hph^3_y+\al^2 g^2\intax{y}\intay{z}\hph^3_y\hph^3_z\nonumber\\
    &\phantom{=}-3i\al g^2 (1+3\al)\intax{y}\intay{z}G_{Ryz}\hph^2_y\hph^2_z\nonumber\\
    &\phantom{=}-6\al g^2 (1+3\al)\intax{y}\intay{z}G_{Ryz}G_{yz}\hph_y\hph_z\nonumber\\
    &\phantom{=}+2i\al g^2 (1+3\al)\intax{y}\intay{z}G_{yz}G_{yz}G_{yz}\;.
\end{align}
For the operator $\heta_g$, we have $\al=-\frac{1}{3}$ and all the loop terms cancel leaving us with
\begin{align}
    \heta_g^{{-1}^\dag}=1+\al g \intax{y}\hph^3_y+\al^2 g^2 \intax{y}\intay{z}\hph^3_y\hph^3_z=T\qty[e^{\al g\intax{y}\hph^3_y}]=\heta_g^{-1} \;.
\end{align}
Hence, at order $g^2$, the operator  $\heta_g^\dag=\heta_g$ is Hermitian.

We can also show that the \comma standard' parity operator 
\be
\hpr(t)=e^{\frac{\pi}{2}\int {\rm d}^3\vec{z}\: \hph(t,\vec{z})\hp(t,-\vec{z})+\hph(t,\vec{z})\hp(t,\vec{z})}
\ee
is $\heta_P$-pseudo-Hermitian. Taking the Hermitian conjugate, we have
\be
\hpr^\dag (t)=e^{\frac{\pi}{2}\int {\rm d}^3\vec{z}\: \hph^\dag(t,\vec{z})\hp^\dag(t,-\vec{z})+\hph^\dag(t,\vec{z})\hp^\dag(t,\vec{z})}\;,
\ee
where we have used the fact that $\hph$ and $\hp$ commute at equal times up to a delta function, giving an over all phase to the operator $\hpr$. 

Using the expansion of $\hph^\dag$ in terms of $\hph$, we can show that $\hp^\dag$ satisfies an analogous pseudo-reality condition, with
\be
\hph^\dag(x)=-\heta_P(t)\hph(x_P)\heta_P^{-1}(t) \qquad\text{and}\qquad \hp^\dag(x)=-\heta_P(t)\hp (x_p)\heta_P(t)\;.
\ee
Hence, using the pseudo-reality condition, we find that
\be
\hpr^\dag(t)=\heta_P(t,t_0)\hpr(t)\heta^{-1}_P (t,t_0)\,,
\ee
implying that the operator $\hpr$ is $\heta_P$-pseudo-Hermitian.

Finally, we can use the above results to show that $\heta_P=\heta_g \hpr$ is Hermitian, viz.,
\be
\heta_P^\dag =\hpr^\dag \heta_g^\dag=\heta_P\hpr\heta_P^{-1}\heta_g=\heta_P \hpr \hpr^{-1}\heta_g^{-1}\heta_g=\heta_P\;,
\ee
as required.




\begin{thebibliography}{99}

\bibitem{Bender:1}
C.~M.~Bender and S.~Boettcher,
``Real spectra in non-Hermitian Hamiltonians having $\mathcal{PT}$ symmetry,''
Phys.\ Rev.\ Lett.\ \textbf{80} (1998) no.24, 5243--5246
[arXiv:physics/9712001 [physics]].

\bibitem{Bender:2}
C.~M.~Bender, S.~Boettcher and P.~Meisinger,
``$\mathcal{PT}$ symmetric quantum mechanics,''
J.\ Math.\ Phys.\ \textbf{40} (1999) no.5, 2201--2229
[arXiv:quant-ph/9809072 [quant-ph]].

\bibitem{Bender:2009mq}
C.~M.~Bender and P.~D.~Mannheim,
``$\mathcal{PT}$ symmetry and necessary and sufficient conditions for the reality of energy eigenvalues,''
Phys. Lett. A \textbf{374} (2010) nos.15--16, 1616--1620
[arXiv:0902.1365 [hep-th]].

\bibitem{Most:1}
A.~Mostafazadeh,
``Pseudo-Hermiticity versus $\mathcal{PT}$ symmetry:\ The necessary condition for the reality of the spectrum of a non-Hermitian Hamiltonian,''
J.\ Math.\ Phys.\ \textbf{43} (2002) no.1, 205--214
[arXiv:math-ph/0107001 [math-ph]].

\bibitem{Most:2}
A.~Mostafazadeh,
``Pseudo-Hermiticity versus $\mathcal{PT}$-symmetry. II. A Complete characterization of non-Hermitian Hamiltonians with a real spectrum,''
J.\ Math.\ Phys.\ \textbf{43} (2002) no.5, 2814--2816
[arXiv:math-ph/0110016 [math-ph]].

\bibitem{Most:3}
A.~Mostafazadeh,
``Pseudo-Hermiticity versus $\mathcal{PT}$-symmetry III:\ Equivalence of pseudo-Hermiticity and the presence of antilinear symmetries,''
J.\ Math.\ Phys.\ \textbf{43} (2002) no.8, 3944--3951
[arXiv:math-ph/0203005 [math-ph]].

\bibitem{Bphi1}
C.~M.~Bender, D.~C.~Brody and H.~F.~Jones,
``Scalar quantum field theory with cubic interaction,''
Phys.\ Rev.\ Lett.\ \textbf{93} (2004) no.25, 251601
[arXiv:hep-th/0402011 [hep-th]].

\bibitem{Bphi2}
C.~M.~Bender, D.~C.~Brody and H.~F.~Jones,
``Extension of $\mathcal{PT}$-symmetric quantum mechanics to quantum field theory with cubic interaction,''
Phys.\ Rev.\ D \textbf{70} (2004) no.2, 025001
[erratum: Phys.\ Rev.\ D \textbf{71} (2005) 049901]
[arXiv:hep-th/0402183 [hep-th]].

\bibitem{Bphi3}
C.~M.~Bender, V.~Branchina and E.~Messina,
``Ordinary versus $\mathcal{PT}$-symmetric $\phi^3$ quantum field theory,''
Phys.\ Rev.\ D \textbf{85} (2012) no.8, 085001
[arXiv:1201.1244 [hep-th]].

\bibitem{Bphi4}
C.~M.~Bender, V.~Branchina and E.~Messina,
``Critical behavior of the $\mathcal{PT}$-symmetric $i\phi^3$ quantum field theory,''
Phys.\ Rev.\ D \textbf{87} (2013) no.8, 085029
[arXiv:1301.6207 [hep-th]].

\bibitem{Sphi1}
A.~M.~Shalaby,
``Non-perturbative calculations for the effective potential of the $\mathcal{PT}$ symmetric and non-Hermitian ($-g \varphi^4$) field theoretical model,''
Eur.\ Phys.\ J.\ C \textbf{50} (2007) 999--1006
[arXiv:hep-th/0610134 [hep-th]].

\bibitem{Sphi2}
A.~Shalaby and S.~S.~Al-Thoyaib,
``Non-perturbative tests for the asymptotic freedom in the $\mathcal{PT}$-symmetric $(-\phi^{4})_{3+1}$ theory,''
Phys.\ Rev.\ D \textbf{82} (2010) no.8, 085013
[arXiv:0901.3919 [hep-th]].

\bibitem{Bphi5}
W.~Y.~Ai, C.~M.~Bender and S.~Sarkar,
``$\mathcal{PT}$-symmetric $-g\varphi^4$ theory,''
Phys.\ Rev.\ D \textbf{106} (2022) no.12, 125016
[arXiv:2209.07897 [hep-th]].

\bibitem{Mannheim:2018dur}
P.~D.~Mannheim,
``Goldstone bosons and the Englert-Brout-Higgs mechanism in non-Hermitian theories,''
Phys.\ Rev.\ D \textbf{99} (2019) no.4, 045006
[arXiv:1808.00437 [hep-th]].

\bibitem{Alexandre:2017foi}
J.~Alexandre, P.~Millington and D.~Seynaeve,
``Symmetries and conservation laws in non-Hermitian field theories,''
Phys. Rev. D \textbf{96} (2017) no.6, 065027
[arXiv:1707.01057 [hep-th]].

\bibitem{Sablevice:2023odu}
E.~Sablevice and P.~Millington,
``Poincar\'e symmetries and representations in pseudo-Hermitian quantum field theory,''
Phys. Rev. D \textbf{109} (2024) no.6, 6
[arXiv:2307.16805 [hep-th]].

\bibitem{Alexandre:2020gah}
J.~Alexandre, J.~Ellis and P.~Millington,
``Discrete spacetime symmetries and particle mixing in non-Hermitian scalar quantum field theories,''
Phys. Rev. D \textbf{102} (2020) no.12, 125030
[arXiv:2006.06656 [hep-th]].

\bibitem{Alexandre:2022uns}
J.~Alexandre, J.~Ellis and P.~Millington,
``Discrete spacetime symmetries, second quantization, and inner products in a non-Hermitian Dirac fermionic field theory,''
Phys. Rev. D \textbf{106} (2022) no.6, 065003
[arXiv:2201.11061 [hep-th]].

\bibitem{Chernodub:2021waz}
M.~N.~Chernodub and P.~Millington,
``IR/UV mixing from local similarity maps of scalar non-Hermitian field theories,''
Phys. Rev. D \textbf{105} (2022) no.7, 076020
[arXiv:2110.05289 [hep-th]].

\bibitem{Bender:2012yv}
C.~M.~Bender and D.~W.~Hook,
``Universal spectral behavior of $x^2(ix)^\epsilon$ potentials,''
Phys. Rev. A \textbf{86} (2012) no.2, 022113
[arXiv:1205.4425 [hep-th]].

\bibitem{Most:4}
A.~Mostafazadeh,
``Pseudo-Hermitian representation of quantum mechanics'',
Int.\ J.\ Geom.\ Meth.\ Mod.\ Phys.\ \textbf{07} (2010) no.07, 1191--1306
[arXiv:0810.5643 [quant-ph]].

\bibitem{Li:2023kpi}
Y.~D.~Li and Q.~Wang,
``PT-symmetric quantum field theory in path integral formalism and arbitrariness problem,''
[arXiv:2305.05809 [hep-th]].

\bibitem{Li:2024xms}
Y.~D.~Li and Q.~Wang,
``Isospectral local Hermitian theory for the $\mathcal{PT}$-symmetric $i\phi^3$ quantum field theory,''
Phys. Rev. D \textbf{111} (2025) no.2, 025016
[arXiv:2412.10732 [hep-th]].

\bibitem{Fring:2019hue}
A.~Fring and T.~Taira,
``Goldstone bosons in different PT-regimes of non-Hermitian scalar quantum field theories,''
Nucl. Phys. B \textbf{950} (2020), 114834
[arXiv:1906.05738 [hep-th]].

\bibitem{Alexandre:2018xyy}
J.~Alexandre, J.~Ellis, P.~Millington and D.~Seynaeve,
``Gauge invariance and the Englert-Brout-Higgs mechanism in non-Hermitian field theories,''
Phys. Rev. D \textbf{99} (2019) no.7, 075024
[arXiv:1808.00944 [hep-th]].

\bibitem{Millington:2019dfn}
P.~Millington,
``Symmetry properties of non-Hermitian $\mathcal{PT}$-symmetric quantum field theories,''
J. Phys.:\ Conf. Ser. \textbf{1586} (2020) no.1, 012001
[arXiv:1903.09602 [hep-th]].

\bibitem{Alexandre:2019jdb}
J.~Alexandre, J.~Ellis, P.~Millington and D.~Seynaeve,
``Spontaneously Breaking Non-Abelian Gauge Symmetry in Non-Hermitian Field Theories,''
Phys. Rev. D \textbf{101} (2020) no.3, 035008
[arXiv:1910.03985 [hep-th]].

\bibitem{Witten:2021nzp}
E.~Witten, ``A note on complex spacetime metrics'', in A.~Niemi, K.~K.~Phua and A.~Shapere (eds.), {\it Frank Wilczek:\ 50 Years of Theoretical Physics}, World Scientific, Singapore, 2022, pp.~245--280
[arXiv:2111.06514 [hep-th]].

\end{thebibliography}
\end{document}